\documentclass[pre,twocolumn,superscriptaddress,showpacs,amssymb]{revtex4}
\usepackage[dvipdfm]{graphicx}
\begin{document}
%            % start of the contribution
%
%\title{Symmetry and the Evolution of Canalization in 
%Highly Connected Networks of Competing Boolean Nodes}
\title{Canalization in the Critical States of
Highly Connected Networks of Competing Boolean Nodes}
\author{Matthew D. \surname{Reichl}}
	\affiliation{Department of Physics, University of Houston, Houston, Texas 77204-5005, USA}
	\affiliation{Texas Center for Superconductivity, University of Houston, Houston, Texas 77204-5002, USA}
\author{Kevin E. \surname{Bassler}}
	\affiliation{Department of Physics, University of Houston, Houston, Texas 77204-5005, USA}
	\affiliation{Texas Center for Superconductivity, University of Houston, Houston, Texas 77204-5002, USA}

\date{\today}


\begin{abstract}        % give a summary of your paper
Canalization is a classic concept in Developmental Biology
that is thought to be an important feature of evolving systems. 
In a Boolean network it is a form of network robustness in which 
a subset of the input signals control the
behavior of a node regardless of the remaining input.
It has been shown that Boolean networks can become
canalized if they evolve through a frustrated competition between nodes. 
This was demonstrated for large networks in which each node had
$K=3$ inputs. Those networks evolve to a critical steady-state at the boarder
of two phases of dynamical behavior. 
Moreover, the evolution of these networks was shown to be associated 
with the symmetry of the evolutionary dynamics. We extend 
these results to the more highly connected $K>3$ cases and show that similar canalized critical steady states emerge with the same associated dynamical
symmetry, but only if the evolutionary dynamics is biased
toward homogeneous Boolean functions.
%                         please supply keywords within your abstract
\end{abstract}
\maketitle 

\section{Introduction}
Boolean networks were originally proposed as models of genetic regulatory networks
and are now widely used as models of self-regulatory
behavior in biological, physical, social, and engineered systems
\cite{Kau69,Kau93,Ald03,Dro08}.
They are designed to capture essential features of the complex dynamics of
real networks by ``coarse-graining'' that assumes that dynamical state
of each node is Boolean, or simply on/off
\cite{Bor05}. 
For example, in a model of a genetic regulatory system each node corresponds to
a gene and its Boolean dynamical state refers to whether 
or not the gene is currently
being expressed.
The regulatory interactions between nodes are described by
a directed graph in which the Boolean (output) state of each node 
is determined by a function of the states 
of the nodes connected to it with directed in-links. 
It has been shown that, despite their simplicity, Boolean networks 
capture many of the 
important features of the dynamics of real self-regulating networks, including
biological genetic 
circuits \cite{Alb03,Li04,Sch05,Lau07}.

Perhaps the most notable feature of Boolean networks is that they have two
distinct phases of dynamical behavior. These two phases are called ``frozen'' 
and ``chaotic'', and in random Boolean networks there is a continuous 
phase transition between them\cite{Der86, Wei86, Fly88}. The two phases 
can be distinguished by how a perturbation in the
network spreads with time: in the frozen phase a perturbation decays with
time, while in the chaotic phase a perturbation grows with time \cite{Der86,Sch04}. 
In networks in which the states of the nodes are updated synchronously, 
the two phases can also be distinguished by the distribution of network's attractor
periods \cite{Ald03, Bas97}. 
When the updates are synchronous the system always settles onto a dynamical attractor of finite
period.
In the frozen phase the distribution of attractor periods is sharply 
peaked with a mean that is independent of the number of nodes $N$.  
In the chaotic phase the distribution of attractor periods is also sharply 
peaked, but with a mean that grows 
as $\exp(N)$.  
In the ``critical'' state, at the boundary between the two phases, the distribution of
attractor periods is broad, described by a power-law \cite{Bha96, Bas97, Pac00, Gre09}.

Many naturally occurring, as well as engineered, self-regulating network 
systems develop through some sort of evolutionary process.
Motivated by this fact,
a number of models that evolve the 
structure and dynamics of Boolean networks have been studied 
\cite{Kau86,Bor98,Pac00,Bor00,Sne00,Luq01,Bas04,Bas05,Liu06,Liu07,Sze07,Roh07,Bra07, Mih09, Sze10, Pri11}. 
These evolutionary Boolean network (EBN) models generally seek to determine 
the properties of networks that result from the evolutionary mechanism 
being considered. For example, some studies have explored evolutionary 
mechanisms that result in networks that have dynamics that are robust again 
various types of perturbations, or that result in networks that are 
in a critical state. 

One example of an EBN is the model of competing Boolean nodes first 
introduced in Ref.~\cite{Pac00}, and later studied in Refs.~\cite{Bas04,Bas05,Liu07}. 
In this model, the Boolean functions of the nodes 
evolve through a frustrated competition for limited resources
between nodes that is a
variant of the Minority game~\cite{Cha97}. 
In the original paper on the model, it 
was shown that the network self-organizes to a nontrivial critical state with 
this evolutionary mechanism. Later it was discovered that this critical state 
is highly canalized \cite{Bas04}. Canalization \cite{Wad42} 
is a type of 
network robustness, and is a classic idea in developmental biology.
Recently, experiments have demonstrated its existence in
genetic regulatory networks \cite{Rut98,Qui02,Ber03}. It occurs when certain 
expression states of only a subset of genes that regulate the expression of a 
particular gene control the expression of that gene.
Canalization is thought to be an important property of developmental biological 
systems because it buffers their evolution, allowing greater underlying 
variation of the genome and its regulatory interactions before some deleterious 
variation can be expressed phenotypically \cite{Wag05}.
Critical networks are thought to important for gene regulation because they can store 
and transfer more information than either frozen networks, which have a 
static output, or chaotic networks, which have a random output \cite{Kau93}.

In the previous studies of the EBN with competing nodes, 
it was found that, 
for large networks in which each node has $K=3$ in-links, 
the system evolves to a steady state
that is both critical and highly canalized.
The canalized nature of the evolved state was found by
measuring the average frequency at which each of the 
256 possible Boolean functions of three variables occur.
It was discovered that the functions organize
into 14 different classes in which all of the functions in each class 
occur with approximately the same frequency. Moreover, all functions in
a class are equally canalizating and  the classes 
whose functions are more canalizing occur with higher frequency. 
It was then found that the existence of these 14 classes is due 
to the symmetry properties of the evolutionary dynamics \cite{Rei07}.

In this paper, we extend these results to more highly connected 
Boolean networks in which nodes have $K>3$ in-links. 
We show that, although an
unbiased implementation of the same evolutionary process
does not lead to a critical state as it does for 
$K=3$, by biasing the 
process in a way that encourages the evolution of more homogeneous
functions the system can evolve to a critical steady state.
We also show that the possible Boolean functions again 
organize into classes that depend on the symmetry of the 
evolutionary dynamics and that all functions in a class occur with 
the same average frequency. In this case however, the bias added to the 
evolutionary process causes a competition 
between the preference for canalization 
and a preference for homogeneity induced by the bias. This competition
between canalization and homogeneity 
is reflected in the frequency at which the Boolean functions occur.

\section{The Model}

\subsection{Random Boolean Networks}

A Random Boolean Network (RBN) consists of $N$ nodes, $i= 1,....,N$, 
each of which have a dynamical state with Boolean value $\sigma_i = $ 0 or 1.
The Boolean state of each node is a function of the Boolean valued states of
a set of $K_i$ other randomly choosen nodes that regulate it. 
The regulatory interactions of the nodes
are, thus, described by a directed graph. 
For 
synchronously updating Boolean networks, 
which are the only ones considered here, 
the states $\sigma_i$ at time $t+1$ is a function of all states of its $K_i$ 
regulatory nodes $\{i_1, i_2,...,i_{K_i}\}$ at time $t$:
\begin{equation}
\sigma_i(t+1) = f_i[\sigma_{i_1}(t), \sigma_{i_2}(t), ..., \sigma_{K_i}(t)]
\end{equation}
The function $f_i$ is a Boolean function of $K_i$ inputs that determines 
the output of node $i$ for all $2^{K_i}$ possible sets of input values. 
In this particular study, we consider RBNs 
in which $K_i$ is fixed for 
all nodes $i$. No self-links, or multiple in-links from the same node are 
allowed in our models.

In RBNs, each of the different functions $f_i$ is chosen randomly.  
We do this
by choosing the $2^K$ outputs of each of the $N$ functions either to be 
`0' with probability $p$ and `1' with probability $1-p$,
or to be `0' with probability $1-p$ and `1' with probability $p$. 
Which of the two cases is used is chosen with equal probability for each
function $f_i$, 
but remains fixed while assigning all of the individual outputs to 
that particular function. By choosing the Boolean functions in
this way, a symmetry between `0's and `1's exists
on average in the network. Note that the effective range of $p$ is only
from 0.5 to 1.

Note that if 
$p=1/2$, the choice of functions is unbiased and each of the $2^{2^K}$ 
functions are equally likely to be chosen. For $p\neq 1/2$ the choice of 
functions is biased toward homogeneity. The homogeneity of a function $f_i$ 
is defined as the probability that it will output a ``0'', or the probability 
that it will receive a ``1'', whichever is larger, assuming that it 
receives random input. The average homogeneity of the network, $P$, 
is the average homogeneity over the functions of all nodes $i$.

The system state $\Sigma$ of the network at time $t$ is given by the array
of Boolean 
values of the states of each node:
\begin{equation}
 \Sigma(t) = \{\sigma_1(t), ..., \sigma_N(t)\}
\end{equation}
Because the dynamics prescribed in Eq.~1 are deterministic, and the space 
of all possible network states is finite (of size $2^N$), all dynamical 
trajectories eventually become periodic. That is, after some possible 
transient behavior, each trajectory will repeat itself after some number of
discrete time steps
$\Gamma$ to form a periodic cycle given by:
\begin{equation}
 \Sigma(t) = \Sigma(t+ \Gamma)
\end{equation}
The periodic trajectory over this cycle is referred to as the ``attractor'' 
of the dynamics, and the minimum $\Gamma$ that satisfies this equation 
is the ``period'' of the attractor.

As mentioned above, two distinct phases of dynamical behavior, ``frozen'' and ``chaotic'', 
exist for RBNs. For networks with uniform 
$K>2$, networks are in the chaotic phase when $p$ is near 0.5 and in the frozen or "fixed" phase when $p$ is near 0 or 1. There is a continuous 
phase transition between these phases at the so called ``edge of chaos.'' The critical value $p_c$ where this transition occurs
satisfies the expression \cite{Der86,Wei86,Fly88}
\begin{equation} 
1= 2Kp_c(1-p_c)
\end{equation}
We will refer to the networks that are critical because they satisfy Eq.~4 by
construction as ``non-evolutionary'' networks, since such networks have no evolutionary dynamics associated with them.

\subsection{Evolutionary Game of Competing Nodes}

The process for evolving the Boolean functions in the network is: 
\begin{enumerate}
\item Start with an RBN constructed with a bias $p$, and choose
 a random initial state $\Sigma(0)$.

\item Update the state of the network using Eq.~1, and determine the 
  attractor of the dynamics. (The attractor can be found using the algorithm discussed
  in the Appendix of Ref.~\onlinecite{Liu06}.)

\item For each update on the attractor, determine which Boolean value is the output state of the majority of the nodes, and give a point to each node that is part of the majority.

\item Determine the node with the largest number of points; that node ``loses''. If two, or more, nodes are tied, pick one at random to be the loser.

\item Replace the function of the losing node with a new randomly chosen 
  Boolean function with bias $p$. 

\item Return to step 2.
\end{enumerate}

The essential features of the game are (1) frustration \cite{Cha97, Art94}, since most nodes lose each time step, (2)
negative reinforcement, since losing behavior is punished, and (3) extremal dynamics \cite{Bak93},
since only the worst performing node’s Boolean function is changed.

%describe the game's essential features more here

Note that previous studies of this evolutionary model
have always replaced the function of the losing node with an unbiased, $p=0.5$ random Boolean function. Thus, the evolutionary game we study here differs from previous work only
at step 5.

If the attractor period is longer than some limiting time, $\Gamma_{\max}$, 
then the score is kept only over that limited time. 
In our simulations, $\Gamma_{\max} = 10^4$ was used. 
Each progression through the game is called an ``epoch''.

In this evolutionary process only the Boolean
functions evolve; the directed network describing the regulatory interactions between
nodes does not change. However, as we will see, canalization effectively removes 
interactions from the network. Thus, as the Boolean functions evolve and canalization 
increases the directed network of regulatory interactions effectively changes \cite{Rei07}.
Thus, effectively, both the structure of the network and the dynamics of the nodes 
simultaneously evolve, making the model effectively a co-evolving
adaptive network model \cite{Bor00, Gro08}.

\section{Canalization and Symmetry}
\subsection{Canalization and Ising Hypercubes}
As mentioned above, for large networks with $K=3$, 
it has been found~\cite{Bas04,Liu07} that the $256=2^{2^K}$ possible Boolean 
functions of three variables organize into 14 different classes in which the 
functions belonging to each class occur with the same frequency in the critical 
steady-state that results from the evolutionary game. Moreover, 
all the functions in a class are equally canalizing, and,
in the steady state, functions
in classes that are more canalizing occur with higher frequency.
Thus, for $K=3$ the game causes networks to evolve to a critical 
steady state that is highly canalized. Canalization occurs in Boolean networks 
when the Boolean functions assigned to the nodes are canalizing. A Boolean 
function is canalizing if its output is fully determined by a specific value of 
one, or more of its inputs, regardless of the value of the other inputs. The 
canalization of a function can be quantified by a set of numbers 
$\textbf{P}_k, k= 0, 1, ..., K-1$, which are defined as the fraction of 
the different possible sets of $k$ input values that are canalizing \cite{Kau93}.

Canalization can be further understood by mapping Boolean functions of $K$ inputs
onto configurations, or ``colorings'', of the $K$-dimensional 
Ising hypercube~\cite{Rei07}. 
The Ising hypercube is a hypercube which has each vertex labeled, or colored, 
either `0' or `1' (``black'' or ``white''). In this representation, each of 
the $2^K$ possible sets of input values corresponds to coordinates on a given 
axis of the hypercube. The color of each vertex represents the output value 
of the function 
for the associated input values. The mapping of a Boolean function of $K$ inputs 
to a configuration of the $K$-dimensional hypercube
is one-to-one.

This representation of Boolean functions as colorings of Ising hypercubes 
facilitates quick recognition of canalizing functions. 
For a $K$-dimensional hypercube, the fractions of canalizing inputs 
$\textbf{P}_k$ of a Boolean functions are the fraction of its $K-k$ dimensional 
hypersurfaces that are homogeneously colored. 

\subsection{Symmetry of Evolutionary Dynamics}
This Ising hypercube representation is particularly helpful
for understanding the role of symmetry 
in the evolutionary dynamics. The fourteen (14) different classes that
were observed 
empirically in the original $K=3$ study were later recognized as 
being those
Ising hypercube colorings that are related by cubic symmetry plus 
parity~\cite{Rei07}. Parity in this case refers to simultaneously inverting the Boolean 
values associated with each vertices. An underlying symmetry 
of the evolutionary dynamics was therefore reflected in this symmetry of 
the evolved steady-state. Furthermore, this symmetry preserves 
the canalization values $\textbf{P}_k$
of functions in 
each class, since neither cubic nor
parity operations change the percentage of homogeneously colored 
hypersurfaces.

In mathematical terms the different classes correspond to the group orbits 
of the ``Zyklenzeiger'' group, which is the hyper-octahedral symmetry 
group $O_n$ (where $n=K$) combined with parity \cite{Rei07}. A group orbit 
is the set of configurations that map into each other through applications 
of a group's symmetry operations. The number of orbits $P_G$ can be calculated 
analytically using P\'{o}lya's theorem \cite{Pol37}:
\begin{equation}
 P_G = \frac{1}{|G|} \sum\limits_{g\in G} |X^g|
\end{equation}
where $G$ is the symmetry group acting on the $K$-dimensional Ising hypercube, 
$|G|$ is the number of operators $g \in G$, $X$ is the set of hypercube 
colorings, $X^g$ is the set of colorings that are left invariant by $g$,
and $|X^g|$ is the size of set $X^g$.

To apply this theorem, one must construct all the operators of the group, 
sum the number of functions left invariant by each operator, and divide by 
the total number of symmetry operators. The hyper-octahedral group $O_n$ 
has $n!2^n$ operations. Including parity operations doubles this number. 

A given symmetry operator $g$ can be written as a permutation of the vertex 
numbers on a given hypercube. As a result, each operator $g$ can be expressed 
in terms of its cycle structure $x_1^{b_1}x_2^{b_2}...x_m^{b_m}$, where $\sum_{i=0}^m ib_i= 2^{K}$. 
This notation indicates that $g$ contains $b_1$ cycles of length 1, $b_2$ cycles 
of length 2, etc. 
The complete cycle representation of the hype-octahedral group
for an arbitray dimension $K$ is given by a known recursion relation \cite{Har63}.
For $K=3$ the complete cycle relation is:
\begin{equation}
 x_1^8 + 13x_2^4 + 8x_1^2x_3^2 + 8x_2x_6 + 6x_1^4x_2^2 + 12x_4^2
\end{equation}
where the coefficients of this polynomial indicate the number operators with 
a particular cycle structure.

Without parity, the number of functions left invariant is equal to $2^{N_c}$, 
where $N_c = \sum_{i=1}^m b_i$ is the total number of cycles in the operator. 
Parity must be treated separately; no functions are left invariant by the 
parity operator with any hyper-octahedral operator containing at least one 
cycle of length 1. Thus there are $2^{N_p}$ functions left invariant for the 
operators which include parity, where $N_p = (1-\Theta(b_1))\sum_{i=1}^m b_i$ 
and $\Theta$ is the Heaviside step function.

Thus, applying P\'{o}lya's theorem to the $K=3$ case, we arrive at
\begin{eqnarray}
P_G = (1/96)((2)^8 + 13(2)^4 + 13(2)^4 +  8(2)^4 + \nonumber \\8(2)^2
 + 8(2)^2 + 6(2)^6 + 12(2)^2 + 12(2)^2) = 14
\end{eqnarray}
This is precisely how many function classes were found empirically in the $K=3$ 
case \cite{Bas04, Rei07}. Similarly, the complete cycle representation for the $K=4$ is
\begin{eqnarray}
x_1^{16} +12x_1^8x_2^4 + 51 x_2^8 + 12 x_1^4 x_2^6 + 32 x_1^4 x_3^4 + \nonumber \\ 48 x_1^2 x_2 x_4^3 + 84 x_4^4 + 96 x_2^2 x_6^2 + 48 x_8^2 
\end{eqnarray}
Using this result, accounting for parity, and applying P\'{o}lya's theorem we
calculate $P_G = 222$ for a 4-dimensional hypercube under rotation plus 
parity symmetry. (Note that an erroneous value of $P_G=238$ was reported in 
Ref.~\cite{Rei07}.) This is how many classes of functions 
should be observed in a critical steady state of $K=4$ networks undergoing 
the evolutionary dynamics. Below we show that 
results from numerical simulations are consistent with this prediction. 
Moreover, the frequencies that these functions occur 
show a preference for canalization.

\section{Results}

\subsection{Critical States of K=4 Networks with Competing Nodes}
We have performed exhaustive simulations of ensembles of networks with $K=4$
playing the game of competing Boolean nodes. All of the simulation results
reported in this subsection 
are for networks with $N=999$ nodes.
Simulations were run for evolutionary processes with
biases of $p=0.5$, $p=0.65$, and 
$p=0.75$. 
The frequency that each of the $2^{2^K}=65536$ different functions occurred 
in the evolved steady state was measured by simulating, 
for each $p$, an ensemble of 
13,000 independent network realizations. 
Each realization was initialized with an 
independent random network with random links and different random functions 
biased with the associated $p$ value. The simulation of each realization was 
run for $10^4$ epochs to allow the network to
reach a steady state. 
At the end the simulation of each realization, 
the functions of each node were recorded and 
then used to calculate the average frequency of each function for the ensemble 
of realizations.
\begin{figure}
\centering
\includegraphics[width=0.45\textwidth]{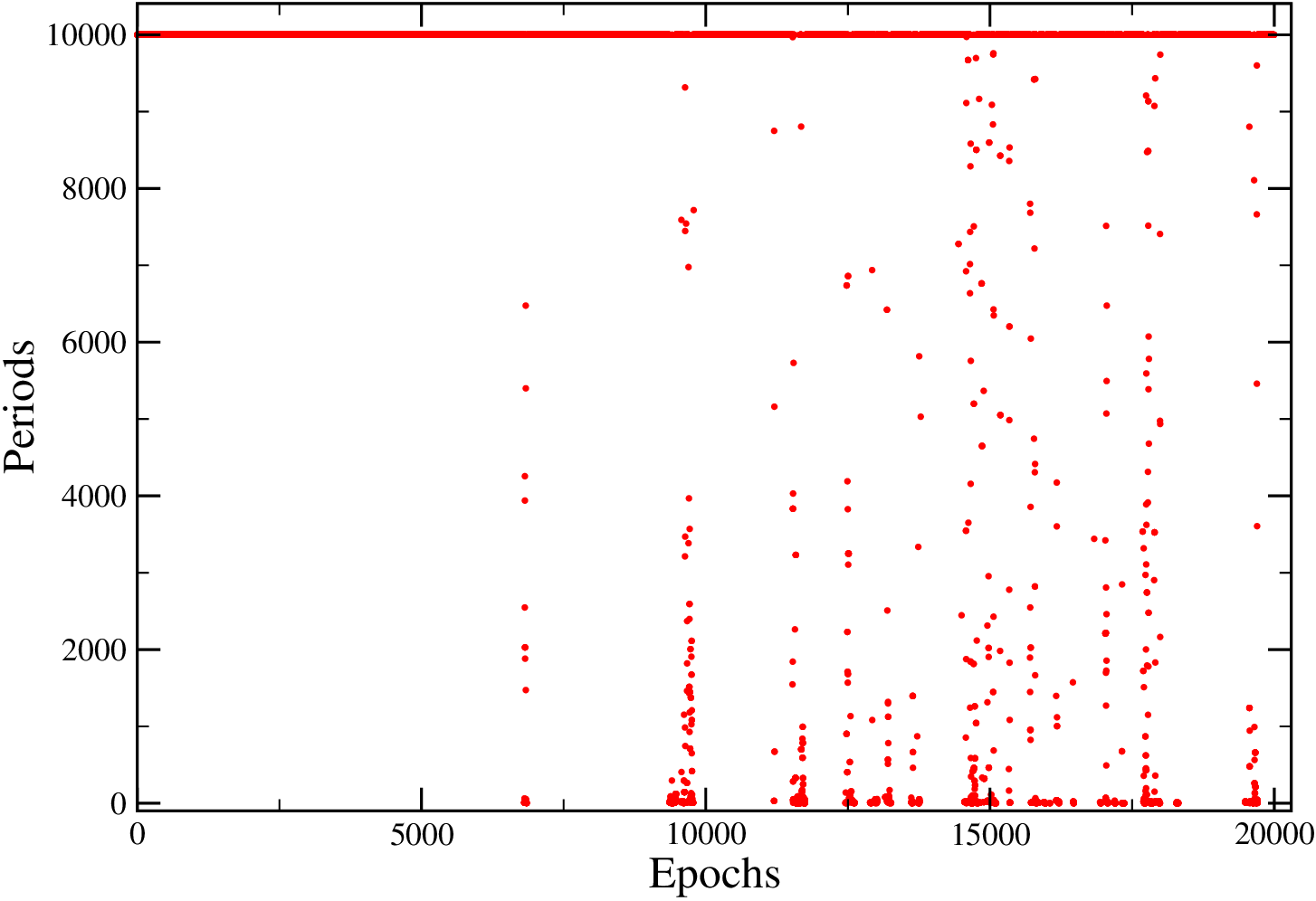}
\caption{(Color online) A graph of attractor period vs. epochs for a
simulation of a $K=4$ network realization with an evolutionary bias of $p=0.65$. Notice that neither short periods nor long 
periods dominate the behavior after the steady state is reached at
$\approx 10^4$ epochs.} 
\end{figure}

Figure~1 shows a graph of the attractor period vs. epoch for a simulation
of a $K=4$ network realization with evolutionary bias $p=0.65$. For 
the first 6834 epochs all the attractors found have periods 
longer than $\Gamma_{\max}=10^4$. Then, attractors with shorter
periods begin to appear. After about $10^4$ epochs
the network reaches a steady state with a broad distribution of attractor
periods. 
As the figure indicates neither 
chaotic behavior, exemplified by almost entirely large attractor periods, 
nor frozen behavior, exemplified by almost entirely short periods, 
dominate the behavior of the steady state. The steady state is instead 
at the ``edge of chaos'' 
and is a critical state. All network realizations 
for $p=0.65$ and $p=0.75$ showed similar behavior once reaching a steady state. 
However, for $p=0.5$, no attractor periods of less than 
10,000 were observed for any 
network realization. That is, unlike in the $K=3$ case, the system does not self-organize 
or evolve 
into a critical state when removing functions and exchanging them with 
\textit{unbiased} functions. This indicates that the increased complexity of 
the more highly connected network disrupts the networks ability to 
self-organize with unbiased functions.

These results indicate that a phase transition occurs in the evolutionary 
dynamics as $p$ is increased. 
At values of $p$ below the transition value the evolutionary
dynamics do not produce a critical steady state, while at values of $p$ above 
the transition, a critical steady state evolves.
In fact, a second transition occurs at higher values of $p$.
At values of $p$ above this second transition, the evolved state is no longer critical
but instead remains in the frozen state. Thus, critical steady states evolve only
when the bias of the evolutionary process is within a range.
%For $K=4$ this transition occurs somewhere below $p=0.65$ (see Figure~4). 

Note that non-evolutionary $K=4$ 
RBNs are constructed with a critical bias of
$p_c\approx0.85355$, according to Eq.~4. This value of $p_c$ produces 
networks with an average homogeneity of all Boolean function in the network
$P \approx 0.85358$. 
However, the steady 
state value of $P$
for networks evolved with a bias of $p=0.65$ 
is $\approx 0.71$ 
Clearly,
the critical state of $K=4$ evolved networks 
is significantly different than the critical state of non-evolutionary $K=4$ RBNs.

\begin{figure}
\centering
\includegraphics[width=0.45\textwidth]{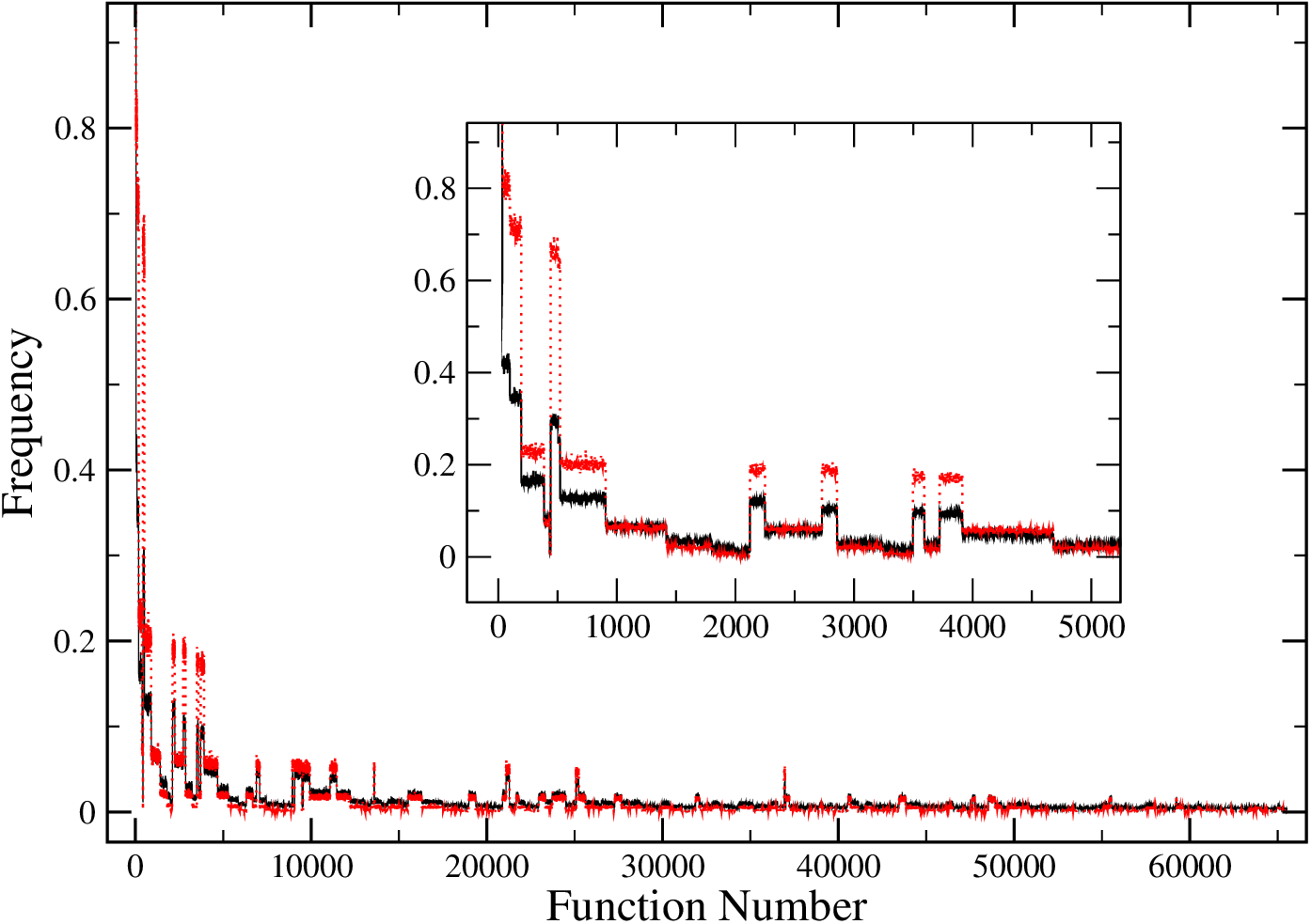}
\caption{(Color online) Ensemble averaged frequency of $K=4$ Boolean functions in the evolved 
critical steady state. The black solid line corresponds to $p=0.65$, and the red dotted line corresponds to $p=0.75$. Functions are 
grouped together in classes based on hyper-octahedral plus parity symmetry 
and with groups then ordered according to canalization from high to low.
The inset shows the same data for functions 1-5000.}
\end{figure}

Figure~2 shows the ensemble averaged frequency at which 
each of the $2^{2^4}=65536$ different $K=4$ Boolean functions occur in the evolved steady 
state, for bias parameters $p=0.65$ and $p=0.75$. The Boolean functions 
were ordered by first grouping them by their membership to a particular  
class under hyper-octahedral plus parity symmetry. Then the classes were ordered
in 
descending order by how canalizing the functions in the class are, as measured by the 
sum of their $\textbf{P}_k$ values. As expected, 
the graph shows that functions belonging to the same 
class occur with the same probability, at least to
within the resolution allowed by statistical fluctuations. 
This confirms the hypothesis that 
the underlying symmetry of the evolutionary dynamics 
is hyper-octahedral plus parity.

Clearly, certain classes of functions occur with a higher probability 
than others. In general, functions on the left side of the graph (higher 
canalization) occur much more frequently than functions on the right 
(lower canalization). However, unlike in the $K=3$ case, certain function 
classes with higher canalization occur less frequently than function classes 
with lower canalization. This occurs because, unlike in the previous
$K=3$ studies
\cite{Pac00,Bas04,Bas05,Liu07}, the evolutionary process here is biased toward 
homogeneity. In this case, the drive for canalization caused by the 
evolutionary dynamics is competing against the bias toward homogeneity for 
certain classes of functions. See Fig.~3. Nonetheless, the evolved critical 
steady state of these biased networks still shows a preference for canalization 
and is strikingly different than a critical state of a non-evolutionary 
RBN where homogeneity 
entirely dominates the relative frequency of functions.

Analogous results presumably hold for even larger values of $K$.
However, because the number of Boolean functions for a given $K$ goes 
as $2^{2^{K}}$, accurately measuring the frequency that each function 
occurs at in the critical steady state
becomes unfeasable for values of $K$ greater than 4.

\begin{figure}
\centering
\includegraphics[width=0.45\textwidth]{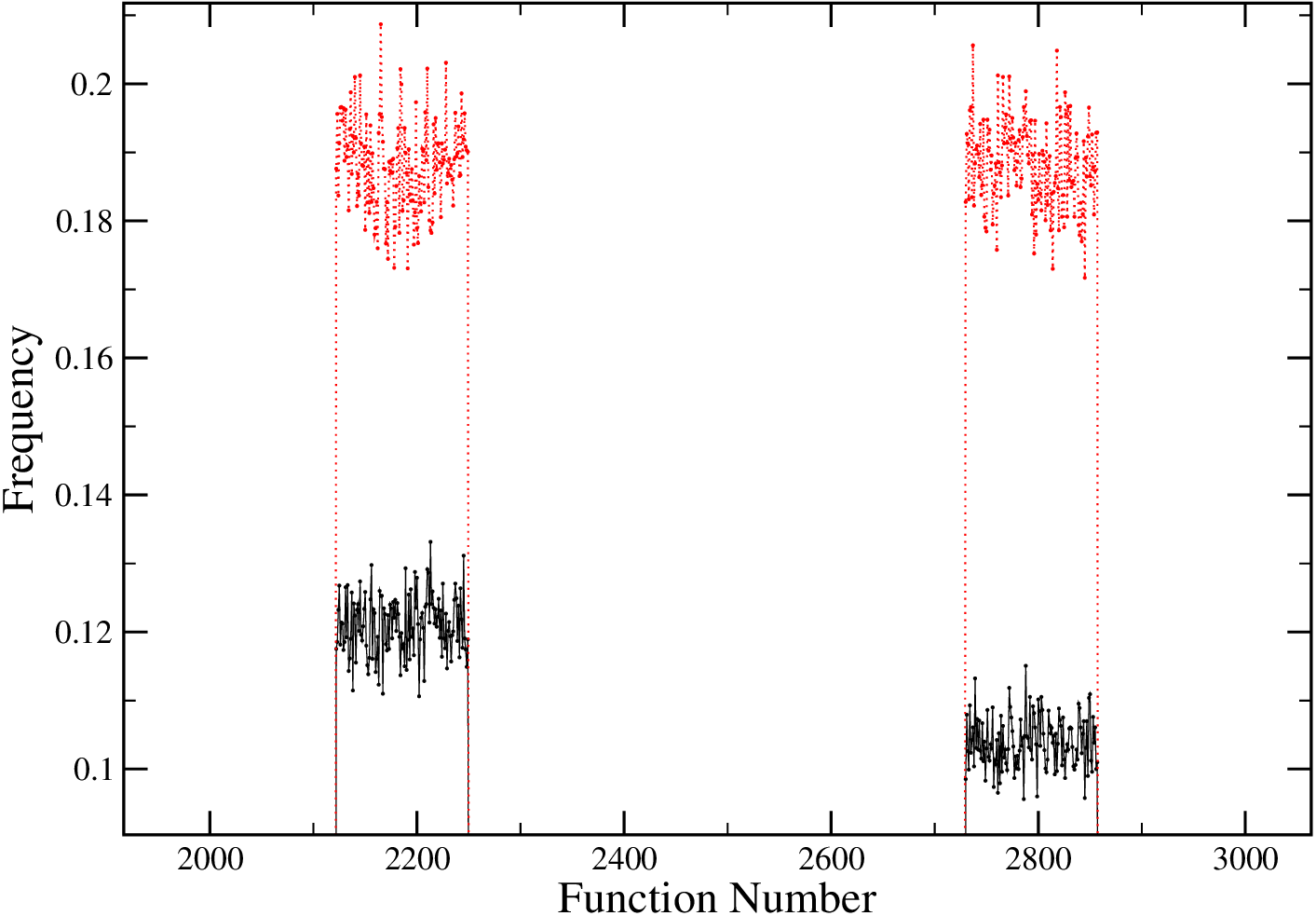}
\caption{(Color online) Enlargement of a small portion of Fig.~2. The two classes of functions 
shown have the same homogeneity but different canalization. 
Notice that for $p=0.65$ (black solid line), 
the class on the left with higher canalization occurs 
with a higher frequency than the class on the right with lower canalization. 
However, for $p=0.75$ (red dotted line) both classes occur with approximately the same frequency.
This indicates that the bias towards homogeneity is dominating the 
behavior over the drive for canalization at the high value of bias for 
these two classes.}
\end{figure}

\subsection{Criticality as a function of $p$ and $K$}
\begin{figure}
\centering
\includegraphics[scale=0.33]{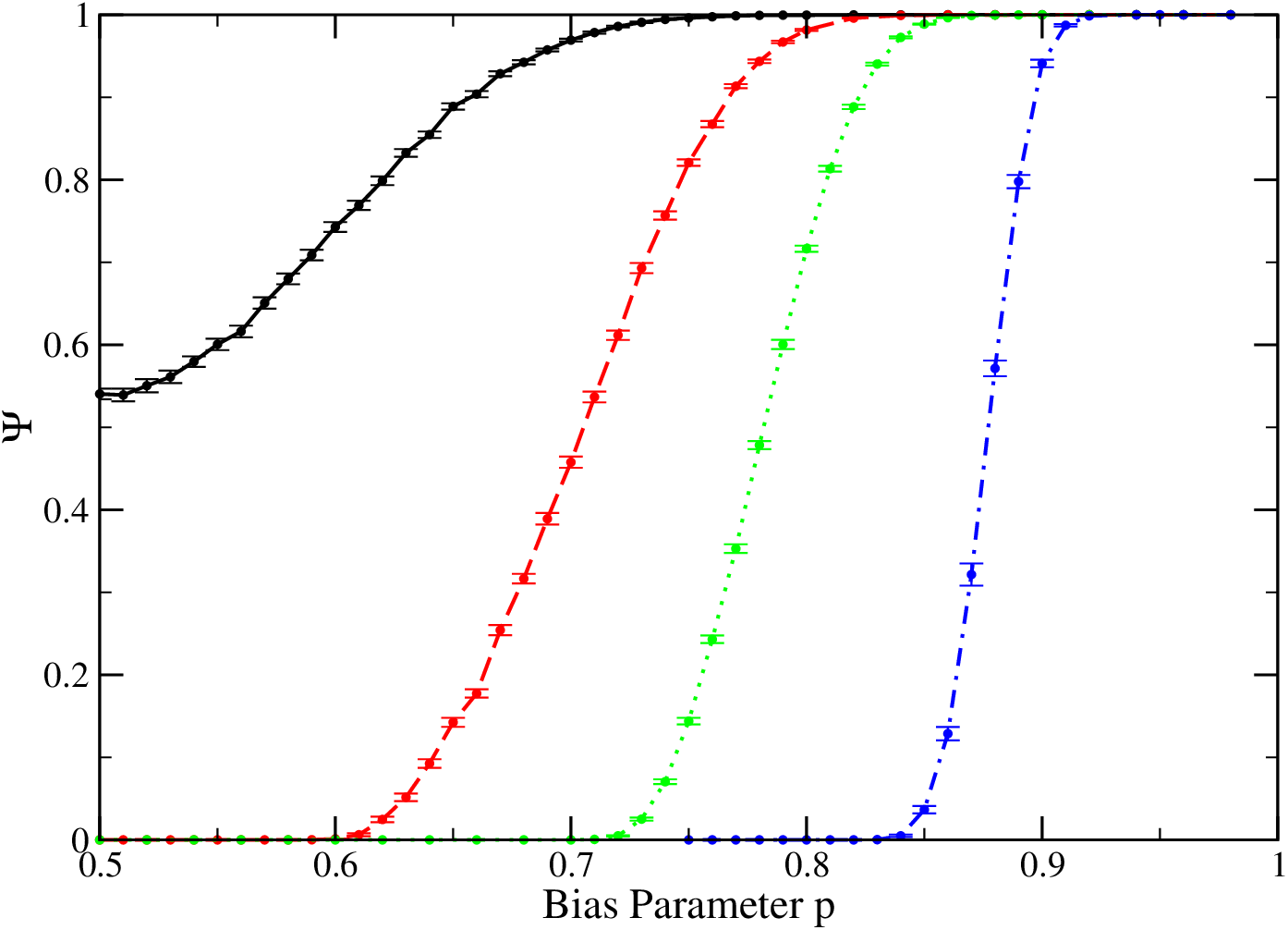}
\caption{(Color online) Critical state evolution order parameter $\Psi$ as a function of 
evolution bias parameter $p$ 
for networks of size $N=999$ with $K=3$ (black straight line), $K=4$ (red dotted line), $K=5$ (green dashed line), and
$K=8$ (blue dashed-dotted line).}
\end{figure}
Our results for $K=4$ indicate that canalized critical states evolve only
when the bias $p$ of the evolutionary game is within a range.
At too low a value of $p$, the evolved steady state is a chaotic state and only
long attractor periods are found. 
At too high a value of $p$, the evolved steady state is a frozen state and only
short attractor periods are found. 
Only in an intermediate range of $p$ does a critical steady state evolve.
In order to quantitatively find the approximate the range of $p$ for which
critical steady states evolve,
we define an order parameter $\Psi$ as the percentage of steady state
attractor periods that occur that are 
less than $\Gamma_{max}$.
Then, when $\Psi = 0$ the networks are assumed to be in the chaotic state, 
when $\Psi=1$ the networks are assumed to be in the frozen state,
and when $0<\Psi<1$ the networks are assumed to be in a critical state. 
%We note that $\Psi$ is measured following an initial number epochs used to equilibrate the network. 

Figure~4 shows a graph of the order parameter $\Psi$ as a function of 
the evolution bias parameter $p$ for network connectivities $K=3,4,5$, and $8$. 
This data was produced using networks of size $N=999$, using an 
equilibration time to reach the steady state of $10^4$ epochs, and then
computing $\Psi$ over $3\times 10^4$ epochs. The $\Psi$ values were also
then averaged over 140 network realizations for each value of $p$ and $K$.
(Sixteen realizations were used for $K=8$). A period cutoff value 
of $\Gamma_{max}= 10^{4}$ was used in these simulations.

For $K=3$ the network already evolves to a critical state at $p=0.5$, the smallest
possible value of $p$, and stops evolving to a critical state  
at $p \approx 0.82$. This is consistent with  
previous results \cite{Pac00} 
that unbiased $K=3$ networks evolve to critical states. 
This is not the case, however, for networks with $K>3$.  As shown in Fig.~4, 
the onset of evolution to criticality for these more highly connected networks occurs at 
a value $p > 0.5$. At least for $K=4,5$, and 8, and presumably for all finite values
of $K$, evolution to a critical state occurs only for a finite range of $p$. For example,
for $K=4$, this range is from $p\approx 0.60$ to $p\approx 0.83$. 
The width of this range appears to decrease, and both the
minimum and median values of $p$ appear to increase, as $K$ increases.
The findings from the results shown in Fig.~4 vary quantitatively, but remain 
qualitatively consistent if
1) the equilibration time is increased, 2) the value of $\Gamma_{max}$ is varied,
or 3) the number of nodes $N$ nodes is changed.

Figure~5 shows the value of $p$ when $\Psi=0.5$, which approximates the median
value of $p$ in the range of the evolution of a critical state,
as a function of $1/K$ for $K=4,5$, and 8. 
Unfortunately, simulating networks with $K\gg 8$ is computationally 
unfeasable with our methods, 
and we are thus restricted to predict the asymptotic behavior of these 
evolutionary random Boolean networks using these relatively small values of $K$.
The three points fall roughly on
a straight line. 
If we extrapolate the linear fit of the data points,
the value of $p$  tends toward a value 
slightly larger than 1 in the limit of large $K$.
However, this is physically unrealizable since $p$ cannot be 
larger than 1. 
Therefore, 
we expect that as 
$K \rightarrow \infty$, the width of the range of $p$ for which criticality occurs 
goes to 0, while the median value of the range goes to 1. 

It is important to note that the range of evolution bias parameter $p$ for which
critical state evolves is, at all studied values of $K$,
less than the critical 
bias value $p_c$ for non-evolutionary RBNs given by Eq.~1.  
Therefore, the critical 
steady state that results from evolutionary process is 
different than the critical
state of non-evolutionary RBNs. From previous studies of 
$K=3$ networks, and from
the results shown in Fig.~2 that were discussed above, 
the difference is that the
evolved critical steady states are more canalized.

\begin{figure}
\centering
\includegraphics[width=0.45\textwidth]{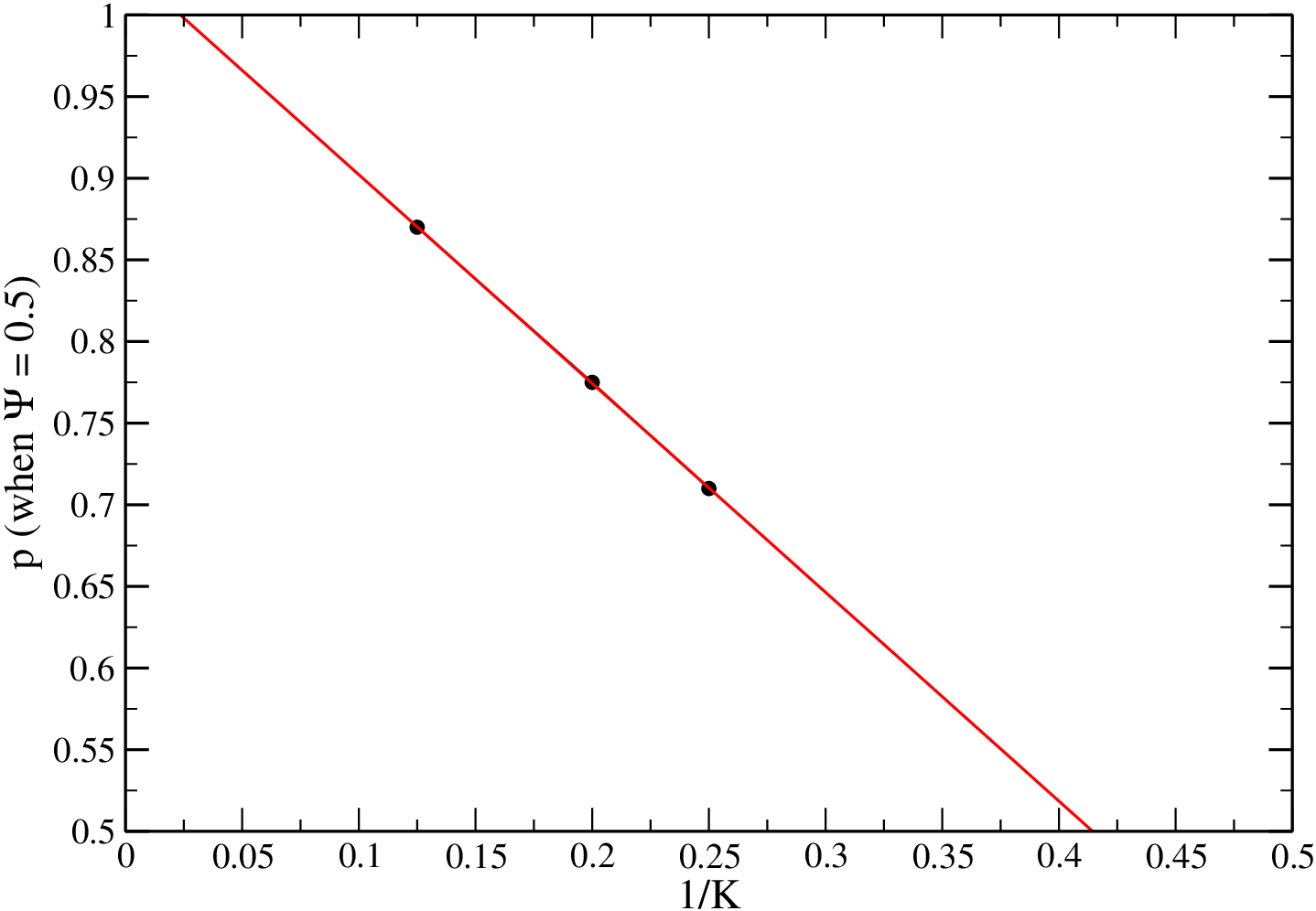}
\caption{(Color online) $p$ when $\Psi=0.5$ as a function of $1/K$ for $K=4,5,8$. The straight
(red) line is a linear fit to these three data points. }
\end{figure}

\section{Discussion}
In this paper we have extended previous work on an EBN model in which the
nodes compete in a frustrated game that causes the Boolean functions of the
nodes to evolve. The previous studies of this model found that the 
game causes the system to evolve to a critical steady state that is 
highly canalized. Canalized states and their evolutionary mechanisms are
important in Developmental Biology beause of the usefulness of the
robustness against diliterous phenotypic expression that canalization
in the genome provides. The previous studies also found that the evolutionary
dynamics of the $K=3$ model has the symmetry of the 3-dimensional Zyklenzeiger
group, which is the combination of parity and the cubic symmetry group.

The previous studies of this EBN, however,
only considered networks with $K=3$ in-links per node. 
Here we extended the study the more highly connected networks with
larger $K$. Real self-regulatory systems, both biological and engineered, 
typically have nodes with a 
range of $K$ inputs \cite{Ald03}. Thus, it is important to
understand how larger regulatory connectivity effects evolutionary
mechanisms. 

For networks with $K>3$, we find that the game as previously studied does not
cause the network to evolve to a critical steady state. Instead, it will evolve
to a chaotic steady state. This occurs because the unbiased game, which was studied
previously, replaces the Boolean function of nodes that lose the game with randomly
selected new functions that are choosen unbiasedly from the set of all possible 
Boolean functions. Apparently, for networks with $K>3$, unlike what happens for 
networks with $K=3$, if they are composed largely of nodes with random Boolean functions 
with an unbiased distribution, then the evolutionary game is not ``strong enough''
to induce a shift in the distribution of the nodes' Boolean functions sufficient
to have critical state dynamics. 

However, we have shown that for networks with $K>3$ if the game replaces the Boolean
function of the losing nodes with functions biased toward homogeneity, then a critical
steady state can evolve. We studied the range of evolutionary bias that will cause
critical state evolution and found that it narrows and that its median increases with $K$.
We have also shown that the critical steady states that evolve for $K>3$ are
highly canalized, although there is also a competing bias toward more homogeneous 
Boolean functions. All functions in an orbit of the $K$-dimensional
Zyklenzeiger group have both equal canalization and equal homogeneity and occur with
equal frequency in the steady state. 
Thus, the symmetry of the evolutionary dynamics of
the EBN with $K$ regulatory links per node is that of the $K$-dimensional Zyklenzeiger
group.

This study illustrates the importance of symmetry in self-regulatory networks and
of evolutionary processes. It would be interesting to analyze other self-regulatory
network systems, both real and model systems, with the methods we have used. This would
allow the importance of symmetry in evolutionary processes to be understood and 
become better appreciated.

%
% ---- Bibliography ----
%

%
\end{document}